\documentclass[12pt, notitlepage]{article}
\usepackage{calc}
\usepackage{natbib}
\usepackage{amsmath}
\usepackage{amscd}
\usepackage{amssymb}
\usepackage{amsfonts}
\usepackage{amsthm}
\usepackage{amsxtra}
\usepackage{euscript}
\usepackage{changebar}
\usepackage{graphicx}
\usepackage{multicol}
\usepackage{epsfig}
\usepackage{psfrag}
\usepackage{setspace} 
\begin{document}
%\begin{frontmatter}
%\title{Statistical Inference for Disordered Sphere Packings}
%\runtitle{Statistical Inference for Disordered Sphere Packings}

%% indicate corresponding author with \corref{}
% \author{\fnms{Jeffrey} \snm{Picka}\corref{}\ead[label=e1]{jdp@math.unb.ca}\thanksref{t1}}
%% \thankstext{t1}{Thanks to somebody} 
% \address{jdp@math.unb.ca \printead{e1}}
% \affiliation{University of New Brunswick}

%%\author{\fnms{???} \snm{???}\ead[label=e1]{???}}
%%\address{\printead{e1}}
%%\and
%%\author{\fnms{???} \snm{???}\ead[label=e2]{???}}
%%\address{\printead{e2}}
%%\affiliation{???}

%\runauthor{J.D. Picka}

%%\begin{abstract}
%%\end{abstract}

%%\begin{keyword}[class=AMS]
%%\kwd[Primary ]{}
%%\kwd{}
%%\kwd[; secondary ]{}
%%\end{keyword}

%%\begin{keyword}
%%\kwd{}
%%\kwd{}
%%\end{keyword}

%\end{frontmatter}

\begin{center}
{\textbf{\Large{Statistical Inference for Disordered Sphere Packings}}}
\vspace{0.1in}
\end{center}

\begin{center}
Jeffrey D. Picka \\
Department of Mathematics and Statistics, University of New Brunswick

\vspace{0.25in}
\textbf{Abstract}
\end{center}

\vspace{0.1in}

Sphere packings are essential to the development of physical models for powders, composite materials, and the atomic structure of the liquid state. There is a strong scientific need to be able to assess the fit of packing models to data, but this is complicated by the lack of formal probabilistic models for packings. Without formal models, simulation algorithms and collections of physical objects must be used as models. Identification of common aspects of different realizations of the same packing process requires the use of new descriptive statistics, many of which have yet to be developed. Model assessment will require the use of large samples of independent and identically distributed realizations, rather than the large single stationary realizations found in conventional spatial statistics. The development of procedures for model assessment will resemble the development of thermodynamic models, and will be based on much exploration and experimentation rather than on extensions of established statistical methods. 

\vspace{0.25 in}

\section{Introduction} 

A disordered sphere packing process can be loosely defined as the type of spatial stochastic process whose realizations describe the final positions of a collection of spherical objects which are tossed into a container and come to rest. Sphere packings are used in physics and engineering to model the atomic structure of amorphous solids, the internal structure of composite materials, and the static and dynamic  properties of powders. They are examples of spatial stochastic processes which need to be fit to data in order to be properly used, yet no formal procedure has yet been developed for objectively assessing the quality of fitted models. These processes have a very different within-realization disorder to that found in processes normally studied in spatial statistics, and so very little study of these processes has been undertaken by statisticians or probabilists. Most work on packings has been undertaken by physicists and engineers who tend to focus on the mean behaviour of the processes and to exclude consideration of their variability, although their work has highlighted a serious need for fit assessment. By combining ideas and methods from spatial statistics and from physics, it is possible to develop a formal approach to inference which is useful in a scientific and engineering context. This approach to inference is also useful in the context of developing methods for fitting spatial models to highly dependent spatial data which is not amenable to fitting by conventional statistical methods. 

It will be generally be assumed that the spheres in packings are of equal diameter (\textit{monodisperse}) rather than being of many different sizes (\textit{polydisperse}). Spheres in $\mathbb{R}^2$ will be termed discs, spheres in $\mathbb{R}^3$ will be termed spheres, and in higher dimensions they will be termed hyperspheres. In the applied literature, planar packings of discs are sometimes referred to as packings of rods are cylinders. This is misleading, since it assumes absolute rigidity for the physical rods and cylinders being modeled.
A \textit{physical packing} is a collection of spherical objects assembled by tossing those objects into a container and allowing a force to form them into a jammed static arrangement. 
%Discs are sometimes referred to as rods in the physics and engineering literature. This is misleading, as any packing of aligned physical rods of great  length would be subject to twisting and have different configurations at different cross-sections. 
Packings will be assumed to be disordered and not based on a point lattice unless otherwise stated. Ordered packings with point lattice structure are of use in the study of geometry in high-dimensional spaces \cite{rogers,grulek,zong:1999} and in the development of efficient coding \cite{cs:1999}, but these applications have no relevance to the modeling of disordered materials. Packings are also of great interest in more abstract mathematical contexts (e.g. the theory of discrete analytical functions \cite{stephenson}), but these uses have no applications to materials either. 

\subsection{Examples of use}  

Sphere packings were first used in attempts to explain the patterns seen in X-ray diffraction studies of liquids and glasses  \cite{bernal:1959,bernal:1964}. Bernal \cite{bernal:1960} and Scott \cite{scott:1960} attempted to model macroscopically what an amorphous mass of densely arranged atoms might look like, using physical packings of monosized steel spheres as their model. More recent work \cite{aswe} has suggested that this approach to modeling may not be feasible, since the packings of spheres are formed by the action of a different set of forces from those found at the atomic scale. If physical models can be shown to explain the patterns seen, then this use of the physical model would be analogous to the use of linear models in other sciences where the use of a simple model can be useful, even if it is partly incorrect. 

Sphere packings are more commonly used in the modeling of granular materials at rest. Granular materials are collections of solid particles surrounded by fluid or gas which form packings at rest, but which flow like liquids or gases when a sufficiently large force is applied. When put into motion, the complex interaction between the grains makes it impossible to model granular flow by means of conventional methods from statistical physics \cite{jaeger:1996,degennes,balls}. Packings are essential components of models for the static and dynamic behaviour of granular materials which are used the study of avalanches, of mudslides, of soil liquification and erosion, and of the flow of powders through pipes and hoppers. 

Sphere packings can be used in the study of composite materials. In the study of metal sintering \cite{sinter}, a packing can represent the initial state of a metal powder before it is compressed and heated to form a solid metal part. Polydisperse packings can be used as models for some some colloids and for concrete. A colloid is a mixture of two immiscible liquids, in which one liquid may tend to form spherical inclusions within the other. For some compositions, the spherical inclusions in a colloid may appear to form a packing \cite{onoda}. Concrete is a composite of rocks of many different sizes held together by a matrix of  cement \cite{minyou}. Sphere-packing-based models for concrete are highly idealized, but may be the only feasible way to abstractly model such a complex material \cite{shiu}. 

Packings can be used as models for porous structures. A packed bed reactor is a large steel vessel filled with solid catalyst-impregnated particles. Liquid reactants flow in at one end of the vessel, and the product emerges from the opposite end. Design of these reactors requires being able to model both the reaction in the pore spaces around the particles and the transfer of heat through the particles. Sphere packings are the simplest form of particle packing that can be used to model the internal structure of these reactors \cite{chu}.

\section{Defining a Packing Process} 

A sphere packing process is a stochastic process whose realizations are arrangements of spheres which closely resemble the arrangements of spherical objects or atoms produced by physical processes in nature. Ideally, it would be possible to define the process mathematically and then to analyze packing processes through mathematical arguments, as can be done for Poisson processes and spatial Markov processes. Insurmountable difficulties posed in defining the packed state and in defining the process make it impossible to undertake an analytical approach to inference. 

The first problem is to define what it means for spheres in a realization to be packed. Consider a configuration of spheres in $\mathbb{R}^d$. Construct the Delaunay triangulation of the sphere centres \cite{obs}. An \textit{interior sphere} is any sphere whose centre has no triangulation edges linked to any of the synthetic points used to construct the triangulation. \textit{Boundary spheres} are spheres which are not interior spheres. Identify all points of contact between spheres and their neighbours, and define the \textit{contact network} to be the subset of the triangulation which joins centres of spheres in contact. If confining surfaces are present, then the contact network must be augmented by edges connecting points of confining surface contact with the corresponding sphere centres. A sphere is \textit{jammed} if it cannot be translated without overlapping another sphere. 
A boundary sphere is jammed if it cannot be translated without overlapping another sphere or penetrating a confining surface. Otherwise, it is a \textit{free boundary sphere}. 

Given an configuration  of spheres, then the simplest definition of a packing would require that all interior spheres be jammed and that the contact network consist of a single connected component. This definition is too strict, in that it is possible to have boundary spheres which are not jammed (called \textit{rattlers}) inside of cages of jammed spheres which may also include confining surfaces. The simple definition is also not strict enough in many applications, in that it may include arrangements of spheres which are unstable when the packing is under the influence of forces. If forces are acting on the packing, then all of the spheres including those on the free boundary need to be in mechanically stable positions. 

Given these definitions, packings of monosized spheres can only be shown to exist by mathematical arguments if the spheres are arranged in a regular lattice structure, such as a hexagonal or cubic close packing. There is no way to demonstrate mathematically the existence of a disordered packing of monosized spheres. Spherical physical objects can be packed by mechanical processes, but the physical objects involved are neither exactly monosized or exactly spherical. Given any list of sphere centre locations in a possible packing, the exact distance between the points can never be exactly known, and so it cannot be said for certain if the list represents a disordered packing of monosized spheres. It is possible that there is no such thing as a disordered packing of monosized spheres, which would complicate the formal definition of a packing process. The same difficulties with existence apply to polydisperse packings, with some exceptions  related to circle packings in the plane \cite{stephenson}.

If disordered packings could be shown to exist, then it would be necessary to formally define a stochastic process which had disordered packings as its realizations. One candidate for such a model is to use the Gibbs process \cite{ruelle} to model the sphere centre positions, which would require that the packing be of infinite extent and be stationary. Gibbs processes are only useful when the only interactions between spheres are pairwise \cite{molwag}. In a Gibbs model for a packing, there would be many higher-order interactions which could neither be ignored nor simply approximated.  

Instead of formally defining a Gibbs model for a packing process, it may be possible to approximate it by explicitly listing the possible configurations for the spheres and then assigning a probability that each configuration will appear in a realization of a packing process. The construction of such configurations would require arguments similar to those used in the identification of the largest fraction of a square (or equilateral triangle or regular hexagon) which can be covered by non-overlapping monosized discs. These calculations can be done analytically for 9 or fewer discs, and computer-based proofs are needed for 10 to 27 spheres \cite{locatelli}. For larger configurations, approximate simulation-based arguments are required \cite{gralub,markot} which makes this approach futile for any reasonably-sized packing.
 
The lack of a useful formally defined model such as the Poisson process or the Gibbs process implies that likelihood-based inference is impossible for any physical packing process. Any inference or model assessment procedure must be based on replications of physical processes or on independent realizations generated by computer programs. 

\section{Useful Models for Packings}

Since mechanically stable packings of physical objects exist in nature, it is scientifically sensible to attempt to represent them by some type of abstract model which is simpler to deal with than the physical objects themselves. While formal mathematical modeling is impossible, other physical systems and complex computer programs can be used as stochastic models instead.

\subsection{Physical Models for Packings}

A physical model for a packing of objects is a different physical system which is easier to replicate and study than the original packing. Examples include using packings of steel spheres to represent the positions of atoms in amorphous solids or to represent the internal structure of catalyst pellets in a packed bed reactor. In the latter case, a scale model of the real system could be prepared out of different materials from the reactor, so that internal images of packed bed could be found. Physical models are also useful for investigating packings of soft objects and objects of irregular shape \cite{ikegami} when these objects are too difficult to simulate. 

The physical model is stochastic because its realizations are generated by a process that can be modeled by a dynamical system which is sensitive to initial and boundary conditions. Pouring spheres from one container to another is an example of this type of physical process, since repetitions of the process under the same general experimental conditions will produce different disordered packed configurations. For any such process, there is a sample space of possible packed object configurations and a density which represents the relative likelihood of each configuration. The goal of model selection in the case of the physical model is to match the sample space and distribution to that of the physical system which is to be modeled. Both of these sample spaces and densities are inaccessible to theoretical argument and difficult to approximate, and so any inference must be based on comparisons of the replications. 

The first major use of physical models involved attempts to model the molecular structure of liquids \cite{bernal:1959}. Scott \cite{scott:1960} and Bernal \cite{bernal:1964,bernal:1960} both used packings of monosized steel spheres  as models for atomic structure. Packings were made inside of thick rubber balloons bound with rubber bands in order to maximize packing density and reduce the effects of gravity. Bernal also filled the balloons with black paint, which was then allowed to set. The agglomerated mass was dissected carefully in order to identify which spheres were in contact, and physical models of the contact network were constructed. More elaborate measurement methods were developed \cite{bernal:1970}, but interest in this approach soon waned on account of the amount of time and effort required to obtain useful results. 

Physical packings are also distinguished by the material that fills the voids around the spheres. A physical process that packs spheres in room temperature air is distinct from one that packs spheres surrounded by warm liquid wax or paint. To make low-density packings, spheres have been packed in a fluid of equivalent density \cite{onoda}. It also possible to make something very much like a packing by combining two immiscible fluids, one of which forms near-spherical inclusions in the other \cite{brujic}. 

Physical packing processes for discs can also be developed. These could be as simple as swirling a tea tray full of thick coins and then inclining the tray until the coins come to rest. Processes of this kind produce realizations which are very close to hexagonal packings, and so more complicated methods are needed to obtain more disordered realizations. Non-overlapping discs can be placed at random locations on a slightly stretched sheet of rubber. The tension in the sheet is released, forcing the discs together into a smaller area. The positions of the discs are recorded, the sheet is retensioned, the discs are placed back on the sheet, and the tension is released again. By cycling through this procedure, a disc packing can be generated \cite{quickenden}. 

Interest in physical packings has been revived as new remote sensing methods have been developed. X-ray tomography can be used to map the interiors of very large packings \cite{aste,aste2,richphil,seidler}. Magnetic resonance imaging \cite{sederman} may be less expensive, but it also produces lower-resolution images and cannot be used on very large packings. Confocal microscopy has been used to study both sphere packings \cite{kohonen,toiya} and colloidal packings \cite{brujic}. 

\subsubsection{What can be learned from physical packings}

Any model of a physical process must obey the constraints on packed configurations which have been revealed by the study of packings of physical objects.

The choice of experimental procedure which defines a physical packing process  has a great impact on the sample space of configurations associated with that procedure. Experiments with monosized spheres poured into containers showed that the density of the initial configuration of the pour could be made more dense by rodding the packing or by vibrating it up and down \cite{scott:1960}. There appeared to be a well-defined limit to the density achievable by these methods, which resulted in the proposition that there was a well-defined physical state termed a dense random packing of spheres. This state has been shown to not be well-defined  \cite{tor:2000}, and it is possible to use mixing procedures in a Couette-Taylor cell \cite{grebenkov} and other methods \cite{pouliquen} to produce denser near-ordered structure packed configurations. The implication for modeling is that an arbitrary packing model cannot be assumed to be sampling from the same sample space of configurations as does the physical process that is being modeled. 

The apparent existence of a well-defined dense random packing opened the question as to whether or not a disordered packing could be denser than a hexagonal close packing of spheres. The mean volume fraction of dense random sphere packings was thought to be  between 0.63 and 0.64, but that did not rule out the existence of some configurations which might have higher volume fractions. For dimensions $2 \leq d \leq 8$, the densest lattice packings can be identified. For discs in the plane, the densest packing of any kind can be analytically proven to be the hexagonal close packing  \cite{cass:1959}. A computer-based proof with the same conclusion has been constructed for spheres \cite{hales}. In dimensions 10 and higher, some non-lattice packings have been found which are denser than any known lattice packings \cite{cs:1999}.

Polydisperse packings are far more disordered than monodisperse packings. One way of disrupting the near-order of physical packings of monosized discs is to randomly introduce a small amount of size variation into the discs. Physical packings of a fixed sphere size distribution can be very difficult to assemble, since the smaller spheres often segregate and accumulate at the base of the specimen. These difficulties also occur in simulated polydisperse sphere packings \cite{yang,rosato}.

\subsection{General Comments on Simulation Models} 

A simulation model is a program which takes output from a random number generator and produces an arrangement of spheres which is close to that of a packing. While these algorithms could in theory be represented by a formal probabilistic model, they are sufficiently complex that no such model could be stated or usefully analyzed. 

\subsubsection{Machine Precision and the Contact Network} 

Any simulated monodisperse sphere packing will be defined by a list of sphere centre locations. With the exception of some cubic packings, most of these numbers will be irrational. In any simulation, sphere location coordinates can only be specified to a finite number of decimal places. It is not possible to determine if the list of points produced by any simulation program form a packing or not. It is generally assumed that any simulated packed configuration can be collapsed to a rigid jammed configuration, but it is not possible to determine exactly which spheres in the collapsed configuration are in contact and which ones are not. Examination of physical packings  shows that they contain many pairs of spheres which are close but not in contact \cite{bernal:1960,finney}. If the simulation is intended to study a physical phenomenon which is very sensitive to the structure of the contact network, then this uncertainly could contribute additional uncertainty to any physical properties estimated from the simulated packing. 

If round-off errors are the only type of error present in the simulation, they can be minimized through use of maximum precision variables in the code. It is also possible to adapt methods used to deal with errors arising from detectors and from Fourier reconstruction in {X}-ray tomographic imaging of physical packings. In those reconstructions, the estimated distance between the closest neighbours was found to be approximately Gaussian \cite{aste}. This distribution was used to estimate the mean number of neighbours in contact, and could be used to define a stochastic or deterministic method for deciding if two spheres are in contact. 

\subsubsection{Boundaries} 

Most physical packings are formed by a confining surface. This may be a plane orthogonal to the direction of the force driving the physical flow that produced the packing, or it may be a completely closed surface that jams every sphere within the packing. Simulations often omit confining surfaces, either to simulate small parts of much larger packings or to avoid difficulties in coding. 

It is possible to avoid using any boundaries. Spheres can be packed on the surface of a hypersphere of one dimension higher \cite{tobochnik}, but this can introduce unwanted effects from hypersphere surface curvature unless many spheres are used. Packings can also be assembled by simulating a force acting towards a central point. It is not clear that a packing simulated by a central force would represent any packing found in nature. 

Confining surfaces can be avoided by imposing \textit{periodic boundary conditions}. Consider a packing in the plane and a rectangle $W$. If the discs are monosized with diameter $\delta$, then any disc whose centre lies within $\delta/2$ of a side of $W$ has the part of the disc outside of $W$ appear on the opposite side of the rectangle. Since this can occur on any edge, the window $W$ is toroidal and the plane is tiled by copies of $W$ containing a disc packing. If the boundary effects vanish after some distance $k \delta$ from the periodic boundary, then the interior of a realization may be indistinguishable from that of a stationary and isotropic packing. Efforts have been made to determine a useful value for $k$ \cite{gotory,taguchi}, but these are often based on a single statistic and may not reflect every effect of periodic boundaries on the structure of a realization. 

Periodic boundary conditions may be used within confining boundaries in order to save computational effort. Within cylindrical confining boundaries, it is possible to pack four periodic cells \cite{osull:2007}. In any small system, this symmetrical structure may have a strong effect on the distribution of statistics calculated from the packing. 

%If hard boundaries are used, then the size of the region can strongly affect the disorder found in a packing.  In small squares in the plane, disc packings resemble distorted cubic lattices rather than the distorted hexagonal lattices seen in larger packings \cite{nurmela}. 
%To obtain hexagonal structures in small regions, it is necessary to use a triangular region \cite{gralub}. 

\subsubsection{Packings of non-spherical objects}

Most simulation models pack monodisperse spheres, since the code for these programs is relatively easy to write. It is dangerous to assume that such models can represent packings of more complicated objects, unless statistical methods can be used to show that simpler models can represent the physical phenomenon of interest. 

Modeling packings of randomly shaped objects is very difficult. 
While a sphere requires only its centre and radius to completely specify it, a reasonable simulation of a small rock might require a long list of numbers to specify the polynomial surfaces that are spliced together to produce a simplified description. When monodisperse spheres are packed, feasible arrangements can be determined by comparing between-centre distances to the diameter of the spheres. For random shapes, very complex code is needed in order to efficiently determine if any overlaps have occurred. Until these problems can be overcome, the best results for random shapes will come from studying physical packings rather than simulated ones. 

It is possible to simulate the packings of simple non-spherical objects. Ellipses can be packed in the plane and ellipsoids can packed in space because these shapes can be easily described and overlaps can be easily determined. Ellipsoids can be shown to pack more densely than spheres  in simulations \cite{buchalter,wilfil,donev:2005b} and in physical experiments \cite{man}. The shapes of ceramic rings have been approximated by assemblies of triangles \cite{nanda:1999} as part of a  simulation of packed beds.  

\subsection{Ballistic Algorithms} 

Ballistic algorithms are the simplest algorithms which can be used to produce configurations of packed spheres or discs. They were the first developed and are relatively simple to code. The algorithms have no basis in physics, and packings generated by these algorithms are less dense than those generated by physical packing processes \cite{jpm}. Ballistic models must be used with caution in physical modeling, as there is no reason to believe that they have the same sample spaces as do physical processes. 

The first ballistic algorithms were developed by Vold \cite{vold:1959b,vold:1960} to model the formation of flocs. Spheres are dropped at random locations onto a surface. They stop falling either when they hit that surface or another sphere. If they hit another sphere, then they either are locked in place with probability $p$, or are rolled down the packed spheres until they reach a stable position. With $p=0$, the algorithm can produce realizations of loose packings. Vold also developed a central version of this algorithm \cite{vold:1964} and one involving rods composed of a set of $k$ spheres in contact with their centres arranged along a line \cite{vold:1959}.

The first high-density central packing simulators built up packings by placing spheres at the closest possible position to the centre of a packing \cite{bennett,adams}. These algorithms produce packings with estimated volume fractions between 0.61 and 0.62. For central disc-packing algorithms, near-ordered patterns can be avoided by seeding the realization with a non-triangular configuration of discs \cite{onaton}. 

Visscher and Bolsterli \cite{vb:1972} developed a non-central  algorithm which added periodic boundary conditions perpendicular to the base. A sphere is  dropped at a random location, and falls vertically until it contacts another sphere. Then, it follows the shortest path along the surface of the already-placed spheres until it comes to rest in a gravitationally stable position. Each sphere drop is repeated $k$ times, and the final position chosen is the position that is closest to the base. The algorithm produces realizations with estimated volume fraction of 0.582 in $\mathbb{R}^3$, which is less dense than a settled physical packing. In the planar version of this algorithm, the initial spheres dropped have a slightly different size in order to prevent the formation of near-ordered realizations.  

The Visscher-Bolsterli algorithm has been extended in various ways. A central version has been developed \cite{rouille}, with the intention of producing packings that contain bridging structures. Non-periodic boundaries have been added to the algorithm, so that the spheres fill a box in space \cite{aparacio} or in the plane \cite{mj1}, or fill a cylinder \cite{soppe:1990,mueller:1997,mueller:2005}. Using periodic boundaries, irregular shapes can be packed \cite{nanda:1999,coelho}. 

The Visscher-Bolsterli algorithm can be modified to pack spheres of different sizes. When the largest and smallest spheres are greatly different in size, the smaller spheres tend to accumulate at the base of the packing unless they are allowed to randomly stick at unstable positions \cite{yang}. More complicated deposition rules have been devised to avoid size segregation in the plane \cite{fho, fho2}. 

The density of physical packings can be increased by shaking them. The density of ballistic packings can be increased by subjecting the initial packing to rearrangements which superficially imitate shaking. 
A simple rearrangement algorithm orders the initially packed spheres by height, then redeposits them at the same planar location using the Visscher-Bolsterli placement rules \cite{jmp2,jmp3}. This algorithm was used to study segregation by size after shaking. In a shaking algorithm for monosized spheres, each sphere in the packing is displaced upwards by a small Normally distributed perturbation, and is then subjected to many small three-dimensional Normal perturbations which are allowed if no collisions occur. Once the number of collisions reaches a set threshold, the packing is collapsed from the bottom using VB deposition rules. This algorithm can increase the estimated mean volume fraction from 0.581 to 0.590, and still results in a loose packing \cite{barker}. If the initial vertical perturbations are not sufficiently small, then the rearrangement can simplify the contact network and decrease the volume fraction. Rearrangement methods can also be applied to packings of irregularly shaped objects in a container with hard boundaries \cite{jia:2001}.

\subsection{Rearrangement Algorithms} 

Algorithms which rearrange the points in point patterns were the first to  achieve volume fractions similar to those seen in physical packings. These programs begin with a realization of a Poisson or a regular point pattern. The points in the pattern are subjected to deterministic or random translations which eventually produce a sphere configuration close to that of a packing. These methods do not attempt to replicate the dynamic interactions between spheres which occur during the formation of a physical packing. 

The Jodrey-Tory algorithm \cite{jt:1981,jt:1985,barg:1991} was the first simulation algorithm to produce realizations with estimated volume fractions similar to those for dense physical packings. It is initialized by a configuration of $n$ points in the interior of a rectangular prism with periodic boundaries. A sphere of unit radius is attached to each point. In the first stage of the process, the radius of each sphere shrinks by 0.0001 units. The distance between each sphere and its nearest neighbour is found, and the closest pair defining overlapping spheres is identified. This pair are moved apart along a line joining the points until the spheres no longer overlap. When the distance between the closest two spheres drops below a threshold, all spheres have their diameter increased by 0.0002 units, and the process repeats. After 2000 cycles, a second routine of shrinking and translation removes all remaining overlaps. Once the initial configuration of points has been chosen, the algorithm is entirely deterministic. There are also versions of the algorithm which pack ellipsoids \cite{bezrukov} and discs of two different sizes \cite{clarkew}. 

The Jodrey-Tory algorithm can be initialized with one of its own previous outcomes. If this done many times, then the configuration  crystallizes to become denser and less disordered \cite{lochmann}. This phenomena is not seen in the simplest physical processes for generating dense physical packings, and it suggests that the Jodrey-Tory algorithm is not sampling from the set all feasible packed  configurations in the same way that a physical packing process does. 

Other deterministic rearrangement procedures use more complicated  rearrangement rules which may depend on all neighbours \cite{speedy:1998} or on fixing the contact network early in the rearrangement \cite{zinchenko}. The latter strategy can be used to maximize disorder in planar disc packings \cite{hinfejo}. 

Rearrangement algorithms can also be inspired by models for the motion of gas molecules, but these models do not represent the physics of the condensation of a gas. Lubachevsky and Stillinger \cite{lubach,lubstil:1990} begin with points uniformly distributed within a region. Random velocities are assigned to each point, and each point begins to grow a disc at a constant rate. When discs collide, the collision is elastic and momentum and energy are conserved. After a few thousand iterations of disc expansion, the discs form a near-packing. There is a three-dimensional version of this algorithm \cite{lubstil:1991} and a version which can pack ellipsoids \cite{donev:2004}. 

Rearrangement rules can also be based on the minimization of  potential functions. One can begin with a uniformly distributed set of sphere centres and then rearrange these centres via a conjugate gradient algorithm which minimizes a potential \cite{ohern}. The potential can be one used in a thermodynamic model of a packing which is stable until a yield stress is exceeded. 

Rearrangement algorithms can also be written with non-periodic boundaries. They can pack spheres on the surface of a large sphere in $\mathbb{R}^4$, avoiding any boundary effects \cite{tobochnik}. They can be used as the basis of programs to estimate the most efficient packing of a small hard-boundaried region by a fixed number of spheres \cite{gralub}. When combined with elements of ballistic algorithms, they can be used to pack cylinders \cite{nolan:1992}.  

\subsection{Dynamic Algorithms} \label{simpow} 

Ballistic and rearrangement algorithms have no basis in the physics of packing formation. Physical packings are generally formed by pouring physical objects into a container or into a pile. In flow, the objects may be nearly packed or barely interacting. When objects strike confining surfaces or other already-packed objects, they expend energy through erosion, through fracturing, through friction, and through deforming themselves and the container. Exactly how this energy expenditure happens during physical packing formation is unknown, as these processes cannot be observed.

It is possible to construct models for the formation of packings which incorporate idealized versions of the unobservable energy expenditure processes. These models, known as Discrete Element Method (DEM) models, were originally developed for modeling rock mechanics \cite{cundall}. DEM models have been proposed for both piling and container-filling processes for both cohesive and cohesionless particles \cite{zhu:2008}. They can be adapted to include processes which happen in viscous fluids and in environments where both viscous fluids and air are present \cite{zhu}. Since they are all based on unverifiable assumptions about interparticle behaviour, it cannot be assumed that any of them will capture the physics of packing formation.

The first DEM packing model was that of Yen and Chaki \cite{yen}, which simulated the formation of a packing in a rigid-walled box. The spheres are initially positioned above the box in a configuration produced by a simple sequential inhibition process \cite{dbg}. As time evolves, gravity acts upon each particle. When particles come into contact, they experience a Herzian force associated with particle deformation and a tangential frictional force. A Van Der Waals force can be introduced to provide a cohesive force between the spheres. When frictional forces are omitted, the algorithm produces realizations of estimated mean volume fraction 0.633 with higher volume fraction variance than expected. When frictional effects are added, the mean volume fraction falls to 0.578. 

Since the initial model of Yen and Chaki, many improvements have been made to DEM models \cite{zhu,zhu:2008}. Rotational interactions have been added \cite{yangzou}, as has the capacity for packing polydisperse spheres \cite{bertrand} and complex shapes \cite{matuttis}. 

%\subsection{Physical Packings of Spherical Objects} 

%There are alternative physical definitions for packings \cite{ohern}. A packing could be defined as a static arrangement of unbound spheres which can be put into motion after application of a sufficiently large stress. This definition assumes that a sufficiently large stress will be experimentally available to induce the flow of spheres, which may not be the case. Since the dynamics of powder flow are not well-understood, it would not be possible to simulate the experiment. 

%
%\subsubsection{Sphere Configurations Related to Packings} %checked

%

%Diffusion-limited agglomerations \cite{evans} are produced by having spheres come into contact through some random mechanism and then stick in place. In the plane, spheres could be dropped at random locations and then fixed at the location where they first hit a boundary or another sphere. The resulting configuration consists of many tree-like chains which may or may not be connected with each other. The configurations are not packings, since the individual spheres are not rigidly immobilized in a jammed state. By slight adjustments of some packing generation algorithms, it is possible to produce a range of different configurations from diffusion-limited agglomerations to loose packings \cite{jpm}. Diffusion-limited agglomerations are used as models for polymer chains and flocs. 

\section{Descriptive Statistics}

Any one realization of a packing from a packing process consists of a long list of object locations, each of which may be accompanied by size, shape, and orientation information. To fit a model to a physical packing process, it is necessary to have statistics which can summarize what each realization from the same model have in common, and which can distinguish between realizations from different models. In physical and engineering applications, it is necessary to have statistics which can characterize the physical response of interest and also to have predictive statistics which explain how the response depends upon packing structure. These statistics may be spatial averages of various kinds, but may also may be measures of variability of extreme behaviour. 

In physical examples, the search for descriptive statistics is analogous to the process which developed the original thermodynamic state variables used in describing ideal gases. Unlike the statistics in classical thermodynamics, descriptive statistics for packings must be thought of as random variables having distributional properties beyond their means which must be modeled. In the physics literature, it is customary to estimate a statistic from a single large sample, and then to equate the estimate to its expected value via an implicit appeal to a law of large numbers. This may be reasonable in some contexts, but for any one statistic there is no simple way to establish how large the packing must be before the between-realization variance of that statistic becomes negligible. 

To be useful for model fitting, descriptive statistics must be able to distinguish between realizations from different models in spite of between-realization variability. Statistics for spatial processes often lack the power to make such distinctions, and so many statistics may have to be tried before a collection can be found that are sufficiently powerful. To find these statistics requires having a library of statistics which can describe packings in as many different ways as can be conceived of. Statistics are needed to describe what the expert eye can see, and also what no human eye can identify. One of the most significant challenges in developing inference for packings is to develop statistics which can be used in these libraries.

\subsection{Random Set Statistics} \label{moments} %checked

Sphere packing processes are examples of random closed sets \cite{math:1975,mol:2005}. The random set $\Phi$  is formally defined by the collection of random indicator functions
\begin{alignat*}{2}
I_{\Phi}(x) &= 1 & \: \: \text{if} \: \: x & \in \Phi \\
&= 0 &  \: \: \text{if} \: \: x &  \not \in \Phi 
\end{alignat*}
for all $x \in \mathbb{R}^d$. Taking the expected value of this indicator function and its products allows moments $M_p$ to be defined at every point $\{ x_1, \ldots, x_p \} \in \mathbb{R}^d$.
\begin{align*}
M_p(x_1, \ldots, x_p)  & = E[I_{\Phi}(x_1)\ldots I_{\Phi}(x_p)] \\ &= Pr[x_1 \in \Phi \: \text{and} \: x_2 \in \Phi   \: \text{and} \ldots \: x_p \in \Phi] 
\end{align*}
These probabilities are defined on ensembles of many realizations, and do not necessarily describe any particular aspects the internal structure of a  single realization. The $k^{th}$ moment is known as the $k-$point correlation function or the $k-$point probability function in statistical physics. 

If the random set is stationary, then reduced moments can be defined:
\begin{align*}
m_p(x_1,\ldots x_{p-1})  & = E[I_{\Phi}(0)I_{\Phi}(x-1)\ldots I_{\Phi}(x_{p-1})] \\ & = Pr[0 \in \Phi  \: \text{and}  \: x_1 \in \Phi  \: \text{and} \ldots \: x_{p-1}    \in \Phi] 
\end{align*}
The reduced first moment $m_1$ is referred to as the volume fraction and $m_2(r)$ as the covariance. 

If $\Phi$ is  also ergodic, then the reduced moments can be estimated from a single realization: 
\begin{align*}
\widehat{m_1} & = \frac{|\phi \cap A|}{|A|} \\
\widehat{m_2} (x) & = \frac{|\phi \cap A \cap \phi_x \cap A_x|}{|A \cap A_x|} 
\end{align*}
where $A$ is the window of observation, $|A|$ is the volume of $A$, and $A_t~=\{~x ~\in ~\mathbb{R}^d~:~x~+~t~\in~A ~\}$.
These estimators are often themselves estimated by sampling at a regular or random arrangement of points within $A$. When the random set is non-stationary, $\widehat{m_1}$ estimates
\begin{equation*}
\frac{1}{|A|} \int_A M_1(t) dt \label{sm1},
\end{equation*}
whose interpretability depends upon the spatial variability of $M_1(t)$. 

%Since random set processes are not Gaussian, the third and higher moments may contain information which shows clear differences between packing processes. Higher moments are defined on inclusion events which are subsets of the ones used to define the first and second moments, and so first and second moment properties may mask the extra information learned from a higher moment. This additional information can be partially unmasked by estimating the reduced third cumulant $\kappa_3$ of the random set, defined by
%\begin{equation*}
%\kappa_3(x,y)=m_3(x,y)-m_1(m_2(x)-m_2(y)-m_2(x-y))+2m_1^3.
%\end{equation*}

Both the packing and the closure of its complement can be considered to be random sets, so long as the boundary between them is well-behaved and not some sort of fractal. This is the case for a sphere packing, and so moments $m^c_k$ of the realization of the complement of the random set can also be defined. Mixed moments can also be defined:
\begin{equation*}
m^{110}(x,y) = Pr[0,x \in \Phi, y \in \Phi_c] = m_2(x)-m_3(x,y).
\end{equation*}
While mixed and complementary moment functions can be calculated from the moments of the random set, they may be of use if they can illustrate differences between processes more accurately than the moments can. 

The most commonly used descriptive statistic for packings is $\widehat{m_1}$. When the random closed packing was believed to be a well-defined physical state, $\widehat{m_1}$ showed that the earliest simulation algorithms were simulating looser packings. Later simulation algorithms were judged by their ability to attain estimated volume fractions close to those of a dense random packing \cite{jt:1985}. This statistic cannot be used alone for characterizing packing processes, since it reveals nothing about interactions between spheres. Local estimates of  $\widehat{m_1}$ can be used to describe differences in structure within non-stationary sphere packings \cite{mueller:1992,nanda:1999}.

Random set second moments $m_2(r)$ are rarely estimated. Instead, point process statistics are used to describe sphere interactions. For monodisperse sphere packings, there is a very close relationship between $m_2(r)$ and the pair correlation function for the sphere centres\cite{ts:1985}.

Moment statistics can also be applied to transformed random sets. If the spheres grow at a constant rate until they fill space, estimates of $m_1$ and the Euler-Poincar{\`e} coefficient can be plotted as a function of the degree of expansion \cite{bhanu,jernot}. 

The complement of the packing is its void structure. This structure can be described by the spherical contact distribution function $S(r)$, which is also known as the pore size distribution function \cite{barker,zinchenko}. Given an arbitrary point $x$ in the void, $S(r)$ is the probability that the nearest point on a sphere to $x$ lies within a distance $r$ of $x$. It can be estimated by finding the distance to the nearest point on a sphere from many locations which have been randomly or systematically sampled from within the void. It has been used to compare physical packings with simulated packings and to investigate the applicability of theoretical approximations for $S(r)$ which arise from statistical mechanics \cite{gotoh}. 

\subsection{Point Process Statistics} 

The centres of each sphere in a packing constitute a realization of a point process. The basic point process statistics are based on the analogues of the first and second moment for a random measure, and on the Palm distribution of the point process \cite{dvj:2003}. 

The intensity $\lambda$, defined for a stationary process to be the mean number of points per unit volume, is seldom estimated since $\lambda$ is a constant multiple of $m_1$. Local estimates of intensity can be used to quantify disorder \cite{tor:2000} and to investigate the internal structure of large physical packings \cite{aste}.  

The $K-$function $K(r)$ is the second reduced moment function of a stationary ergodic point process. The quantity $\lambda K(r)$ is the expected number of points in a disc of radius $r$ about a randomly chosen point, which is not counted in the expectation. The $K-$function is seldom used in the study of packings, and the pair correlation function $g(r)$ is used instead. It is defined by 
\begin{equation*}
g(r)= \frac{1}{db_dr^{d-1}}\frac{\partial K(r)}{\partial r}, \label{g2}
\end{equation*}
where $b_d$ is the dimension of the unit ball in $\mathbb{R}^d$. The pair correlation function may be also be referred to as the radial distribution function, although that name is also applied to the quantity $RDF(r)=\lambda d b_d r^{d-1} g(r)$. The $K$-function has also been generalized to a function defined on the number of $r-$close triples \cite{schladitz} in a realization. Estimation of the $K$-function for stationary ergodic point processes requires careful treatment of window edge effects \cite{ripl:1988}. 

If the point process is not stationary, estimators of $g_2(r)$ for a stationary ergodic process are difficult to interpret. Estimates are made by taking each point on the realization and creating a sequence of thick shells around it. For each shell, the number of other points in the shell is counted. These counts are totaled for each shell type and rescaled based on the shell volume. A histogram of the rescaled totals  is plotted. If the packed spheres have unit radius, a plot of this estimate has a maximum at $r=1$, then has a pair of very sharp local maxima at at $r=1.73$ and $r=2$ \cite{finney}. This type of plot can be used to show the transition from a packing to a looser structure in a DEM model as the Van Der Waals force increases \cite{yangzou}. For planar packings, the estimated pair correlation function can be used to create a pattern that would be seen if X-ray crystallographic methods were applied to a real material with the same structure as the realization \cite{buchalter}. Those familiar with crystallography may find these patterns easier to interpret than estimates of $g_2(r)$.  

Given an arbitrary location $x \in \mathbb{R}^d$, the empty space function $H_s(r)$ is the probability that the nearest point in a realization to $x$ lies within a sphere of radius $r$ centered at $x$. For a stationary point process, the nearest-neighbour function $D(r)$ is the probability that the nearest neighbour to any point in the realization lies within a distance $r$ of that point.  For the Poisson process, these two probabilities are identical. For all other point processes, they can be compared by the J-function \cite{vanl}, defined as
\begin{equation*}
J(r)=\frac{1-D(r)}{1-H_s(r)}.
\end{equation*}
Nearest-neighbour functions can also be extended to the 2nd-nearest-neighbour, the 3rd-nearest neighbour, and so on \cite{hinfejo}. Estimation of all of these quantities require careful attention to window edge effects. 

\subsection{Statistics Based on Triangulations and Tessellations} 

The triangulation of the sphere centres in a packing forms a simple and natural description of packing structure. Near neighbours can be clearly defined as spheres whose centres are connected by a triangulation edge, and the contact network of the packing is a natural subgraph of the triangulation. The dual graph of the triangulation is a simplified description of the void structure of the packing. Statistics based on triangulations and tessellations were initially applied to packings by physicists who were using physical packings as models for the molecular structure of liquids and amorphous solids \cite{bernal:1960,finney}. 

%Statistics of this kind can also be developed for polydisperse packings \cite{lochmann:2007}. 

The Delaunay triangulation and the Voronoi tessellation are generally used as bases for statistics. The tessellation is constructed by finding all points in $\mathbb{R}^d$ which are equidistant between sphere centres. Euclidean distance is used in the construction, but other distance measures can be used to generate different tessellation structures \cite{obs}. The triangulation is the dual of the tessellation, generated by joining pairs of sphere centres which define a tessellation edge. 

The simplest statistics that can be extracted from triangulations and tessellations are lists of characteristics for each cell. For both types of cells, the area, the perimeter, the largest and smallest angles, and the longest and shortest sides can be found. The coordination number of each sphere is defined to be the number of triangulation edges which belong to the contact network. The lists of observations are traditionally summarized by histograms or by summary statistics (mean, standard deviation, minimum, and maximum). In the physics literature, the histograms are often referred to as plots of the distribution of the statistic. This statement is misleading, since the data is a collection of spatially dependent observations. In these cases, the histograms summarize aspects of the structure of an individual realization rather than estimating a population parameter. 
For Voronoi cell areas, Gamma densities have been found to summarize the shape of the histograms \cite{hinfejo,astedim}. 

%There is no way to be certain that any two spheres in a packing are in contact. For physical packings, there are errors of measurement associated with both remote sensing and paint-based methods. Simulation programs produce configurations which are close to being packings, but it is impossible to make final decisions on account of round-off errors and other failings of the algorithms. 
% There is no way to determine if all of the near-contacts would become contacts if the near-packing were to be relaxed into a packed state.  A histogram of nearest-neighbour distances can be plotted, and used to define a threshold below which spheres will be assumed to be in contact. This is an \textit{ad hoc} solution, and may lead to errors in the structure of the network. In spite of this uncertainty, it has been shown that the mean number of spheres in contact for disordered physical packings is much less than for a hexagonal close packing \cite{finney}.

Tessellation statistics have been used to investigate the differences between realizations of different physical and simulated packings. 
Differences between packings are generally specified first in terms of differences in the estimated volume fraction of spheres $\widehat{m_1}$. Plots can be made of mean contact number versus $\widehat{m_1}$ \cite{tass,aste}, or of the standard deviations of Voronoi cell face area and volume against $\widehat{m_1}$ \cite{jjcq}. Comparisons of histograms of contact numbers from different algorithms are not powerful enough to clearly distinguish realizations from different algorithms \cite{jpm}.  

Triangulation-based statistics can also be used to identify possible rattlers.  In a physical packing, there are often a small number of spheres which are not immobilized. The packing surrounding the rattlers consists entirely of jammed spheres forming arch structures, also known as a bridging structures. These structures cause blockages when silos filled with bulk solids are emptied \cite{fowglas,stainforth,to:2001}. The fraction of rattlers in a packing can be used as a measure of packing efficiency \cite{speedy:1998}. Identification of rattlers is complicated by uncertainty about whether or not neighbouring spheres are in contact. 

More elaborate statistics can be developed from the tessellation and the triangulation. The local density of a packing can be defined as the ratio of the sphere volume to the volume of its Voronoi cell \cite{aste}. The escape fraction statistic is the empirical cumulative distribution function of measurements of the largest sphere that could escape from each tessellation cell through the gaps between neighbouring spheres. It has been shown to distinguish between physical packings of different volume fraction \cite{aste}. A topological density has been defined which is based on a notion of topological distance. This distance is defined by choosing a sphere, and then identifying a sequence of shells radiating out from it. The first shell contains only those spheres which touch the central sphere, the second contains all spheres in contact with the first shell but not contained within the first shell, and so on \cite{brunner}. The density is defined as the leading coefficient of the quadratic fit of the number of spheres in each shell to the shell number \cite{okeeffe}. The smallest values of the density are those of point lattices. 

\subsection{Statistics Based upon Measures of Order and Disorder} %checked

When physical packings were first proposed as models for the molecular structure of liquids, researchers sought to determine whether or not packings possessed some form of local order \cite{bernal:1959}. This local order would take the form of small subunits with near-lattice structure, combined in some complicated way to produce the general disorder of the packing. 
 
The presence of an ordered structure in point patterns may not be obvious to the eye. Materials have been found in nature which are quasi-crystalline \cite{sbgc,senechal}. The arrangement of atoms in these materials would appear disordered to the eye, but x-ray diffraction reveals that their structure can be modeled by a projection into $\mathbb{R}^3$ of a higher dimensional regular point lattice. 

Disorder can be defined as the absence of a regular point lattice structure for the sphere centres. The definition of disorder is not constructive, but instead merely identifies all configurations of spheres which are not point lattice structures. While it may be possible to define statistics which identify the degree to which a particular configuration is distinct from a regular lattice, there will be many different ways of measuring these distinctions and none of them will be clearly better than any other. 

It is necessary to distinguish between topological and geometric disorder. A packing is topologically ordered if the sphere centres can be continuously translated so as to transform its Delaunay triangulation into a point lattice without breaking any bonds \cite{ziman}. In a planar packing, topological defects of the lattice structure can be easily identified and counted \cite{mj2}. 

A packing is geometrically disordered if any aspect of the arrangement of the spheres differs from that of a point lattice. If the spheres in a point lattice are reduced in size by a small amount and then randomly perturbed over a small distance, then topological order can be maintained in a geometrically disordered packing. Statistics which quantify geometric disorder are based on local measurements which describe deviations from expected properties of point lattices. 

Statistics can be derived from the locations of contact points on individual spheres, expressed in spherical coordinates. The fourth- and sixth-order spherical harmonic functions can be evaluated for the contact points on each sphere, and then averaged either over the individual spheres or over many spheres. The fourth-order harmonics have non-zero averages in the presence of local cubic lattice structure, while the sixth-order harmonics have non-zero averages in the presence of local  hexagonal close-packed lattice structure. Averages over single spheres have been used to study the structure of very large physical packings \cite{aste}. Averages over many spheres were originally developed to study the emergence of crystallization in simulated liquids \cite{ronchetti,steinhardt}, and improved averages were used to compare realizations of simulated sphere packings \cite{tor:2000}. Averages over many spheres based on the sixth harmonic were used to describe the reduction in disorder observed in simulated sphere packings that had been cycled through the Jodrey-Tory packing simulation algorithm\cite{richard} . Two different many-sphere  averages of the sixth harmonic were used to compare realizations from three different rearrangement models \cite{kansal}. Neither average was powerful enough to distinguish between realizations from different models.   

Statistics based on spherical harmonics cannot identify ordered structure that is found within small clusters of neighbouring spheres.  Statistics which reveal this type of ordered structure can be constructed using the side lengths of Delaunay simplices, which are tetrahedra formed by four Delaunay triangles which share common edges \cite{anikeenko:2006}. In studies of the reduction of disorder over time in Jodrey-Tory packings, measures of  tetrahedracity and quadroctahredracity were found to be more powerful at tracking changes than were were the averages of spherical harmonics \cite{lochmann}. 
 
Other measures of disorder can be found by estimating means and variances of local estimates of intensity \cite{torquato:2000,torstil:2003}. 
%A statistic based on variances can be defined by taking sampling spheres of radius $R$ at locations chosen from a uniform distribution over the realization. Within each sampling sphere, the number of sphere centres is counted. The sample variance $s^2(R)$ of the counts is found for several different values of $R$. Then, $\log(s^2(R))$ is regressed against $R/D$, where $D$ is sphere diameter for packings and the mean nearest-neighbour distance for non-packings. The expected slope of the line will be $d$ for a Poisson process in $\mathbb{R}^d$, but if the slope is closer to $d-1$ then the pattern is defined to be hyperuniform.  One measure of disorder for hyperuniform patterns is 
%\begin{equation*}
%\bar{\Lambda}=\frac{1}{m} \sum_{i=1}^m \frac{S^2(r_i)}{(r_i/D)^{d-1}}.
%\end{equation*}
%Asymptotic values of $\bar{\Lambda}/\widehat{m_1}^{(d-1)/d}$ can be found for ordered patterns and for Kagome lattices. The lowest of these values are found for point lattices. 

\subsection{Statistics Based on Models for Physical Properties} 

Mathematical models for physical phenomena can be applied out of physical context to yield new statistics. If a composite of two electrically insulating materials can be simulated by a sphere packing, then one or both of its phases can be assumed to be conducting. The bulk resistance of the material can be estimated if the internal structure is known. This resistance has no physical meaning, but is based on the solution to a set of partial differential equations which use the packing structure as part of their boundary conditions.  

Statistics based on models for physical properties are traditionally used as response statistics in physical applications. When searching for statistics to explain the response, statistics based on physical models used out of context can be proposed as potential predictors of the response. 

\subsubsection{Statistics Based On Frictionless Flow} 

In frictionless flow, the flow experiences no internal resistance due to shear. Heat flow by means of conductance and the flow of electricity are examples. Models for frictionless flow are relatively simple to construct.  

For electrical flow, the packing and its complement can be considered to be two materials with differing electrical conductivity. On the boundary of the packing, two disjoint sets of spheres can be considered to be connected to electrodes of infinite conductance. When a unit potential is placed between these electrodes, the potential can be calculated at all points within the packing and its complement. From this, a bulk resistance for the composite can be calculated. This resistance defines a mean distance across the packing \cite{klein:1993}, whose form is determined both by the relative resistivity of the two phases and by the disordered structure of the packing.   

For a composite modeled by a disordered packing, the bulk resistance is difficult to calculate. There are no exact methods, and numerical methods require that the packing be discretized very crudely.  If equal resistance is assigned to all of the edges of the contact network, then the potential at all vertices of the circuit can be found easily be means of the properties of random walks through the network \cite{ds:1984}. 
The bulk resistance is a weighted average of potentials at the electrode regions. If the network lies in the plane, the potentials can be plotted and used as a diagnostic tool for finding problems with the potential-calculating software. A plot of the current along each edge, calculated from potential differences between the defining vertices, is more useful than the potential as a descriptor of structure. 
Bulk resistance has been used to develop a test for the presence of anisotropy in sphere packing realizations \cite{picka:2005}. 

Models for heat conduction can also be used. The spheres in a packing can be expanded in order to generate contact surfaces between neighbouring spheres. A plot of the bulk heat conductance of a packing as a function of the degree of spherical expansion has been used to distinguish between packings generated by different models \cite{argento}. 

\subsubsection{Statistics based on shearing flows} \label{flow}

If a fluid flows through a fixed packing, it develops internal frictional losses which depend on the structure through which the fluid flows. If the packing itself is made to flow as a powder, then its flow is strongly affected by frictional losses arising from colliding spheres. These losses greatly complicate the modeling of the flows. 

If the packing is considered to be fixed, the flow of a fluid through its complement can be modeled. All fluid flow is assumed to be laminar, since modeling turbulent flow through a complex structure is impractical. Major simplifications are required in order to be able to calculate bulk flow properties. The complement can be represented by a piping network whose structure is determined from the Voronoi tessellation, and pipe resistances can be assigned on the basis of local void geometry \cite{chu}. The flow through the void structure can also be modeled by a lattice-Boltzmann model \cite{chen:1991}. The velocity profile and pressure gradient of the simulated flow can be used as statistics. If the diffusion of contaminants through the flow is modeled \cite{maier:2003}, many different statistics related to contaminant flux and concentration can be calculated.  

If the packing itself flows, this flow can take place in a vacuum, in air, in a liquid, or in both air and liquid. These flows can be modeled using DEM models which represent compaction processes that do not induce particle fracture \cite{coube,chooi}, flow in mixers and drums \cite{chaudhuri,portillo,sato}, and flow during avalanches \cite{chang:1992,brewster}. Flow in mutiaxial \cite{ngd,david,osull:2007}, and shear test apparatus \cite{thorz,hartl} can also be simulated. In some cases, the simulation program would impose conditions on the structure of the sample. The best statistics would emerge from models for flows which are nearly packed. Flows of fluidized packings would be less useful, since the packed structure would almost immediately be destroyed by the fluidization process. 

Use of DEM models is complicated by the complexity of modeling nearly packed states in motion. In the applications literature, the
unverifiable assumptions made in order to obtain simulation results are not often clearly stated. Modeling some aspects of packing formation remain primitive. In physical powders, some energy is expended through deformation of the grains. Modeling of this deformation can only be undertaken in the plane by means of finite element methods, and only for a small number of grains \cite{zavaliangos,procopio}. More idealized models can be used in model assessment. A plot of the number of sliding contacts as a function of the total number of contacts during simulated flow can distinguish between realizations of different packing models \cite{mcnamara}. 

\subsubsection{Thermodynamic Statistics} 

Given a packing of spheres, each sphere can be very slightly shrunken in place. The shrunken spheres can be acted upon by a pre-specified potential, and the motion of the particles under that potential can be tracked. The motion is confined to a region with periodic boundaries.  Algorithms for simulating these motions have been developed as part of thermodynamic models for hard-sphere atoms \cite{gruhn,gruhnmon,donev:2005a}, and also for non-spherical shapes \cite{donev:2005b}. Once these algorithms have been run for some time, statistics representing pressure, temperature, and other thermodynamic quantities can be calculated. 

\subsection{Graphical Methods} 

New statistics can be developed through quantification of characteristic features of patterns seen in plots of packings. 
A simple plot of the discs in a realization can reveal the near-ordered arrangements of some planar packings. In space, cross-sections showing features of structure can be made from tomographic data \cite{aste}. Two-dimensional projections of tomographic data from physical sphere packings in cylinders reveal the ordered structure of the packing near the cylinder wall, in contrast to the disordered structure near the centre \cite{seidler}. 

In the planar case, extra features can be added to images of the packing. The Delaunay triangulation can be plotted on top of the discs, and triangles can be coloured to identify triples of discs in mutual contact \cite{lubstil:1991}. Edges of the triangulation which are associated with defects can be coloured to illustrate the overall structure of those defects in large packings \cite{mj1}. 

\subsection{Other Statistics} 

There is no evidence that statistics based on presently existing physical and statistical theory will be powerful enough to distinguish between the realizations of a given pair of packing processes.

Experimental physics is an important source of new statistics. Size segregation in agitated granular matter is a phenomenon that would not be predicted from theory or from simple intuition. It occurs when physical packings of particles of differing sizes or densities are shaken \cite{williams,bridgwater}. An example of this is the Brazil nut paradox, in which the large and heavy Brazil nuts rise above smaller and lighter nuts when the container of nuts is shaken \cite{rosato}. Simulated sphere packings have been used to investigate the paradox \cite{rosato2,hong}. These simulation experiments could be used as sources of new statistics, particularly in cases where there is a size distribution among the spheres. 

\subsection{Smoothly Defined and Structurally Defined Statistics}

A statistic is \textit{smoothly defined} if it is based on measurements taken at many locations in the realization, and if the calculation of the statistic treats these measurements as if they arose from a random sample. The observation locations can be taken on a fine grid or from a realization of a Poisson process, but their definition is completely independent of the structure of the packing. To estimate a stationary 2-point correlation function at the vector $r$, a fine grid of sampling locations and its translate by $r$ are imposed upon the realization. For each sampling location and its translate, a 1 is assigned if both points are in spheres, else a 0 is assigned. The estimator is the average of the measurements. While this statistic is considered to be a simple measure of interaction, its construction combines information about voids and neighbouring particles with no consideration of the overall structure of the realization. 

Other smoothly defined statistics can be constructed from the Delaunay triangulation or the Voronoi tessellation. When a histogram is made of all triangle areas from a triangulation, all information about how the triangles fit together is lost. The estimation of the nearest neighbour function at a distance $r$ is done by finding the length of the shortest triangulation edge from each vertex, and assigning a 1 if it is less than $r$ in length and a 0 otherwise. The estimator is again an average, and all other information about structure is ignored. 

Statistics based on physical properties differ from the smoothly averaging statistics in that their construction depends on the disordered structure of the entire realization. Bulk electrical resistance can be thought of as a distance across the network on which current flows \cite{klein:1993}. This distance is based on a weighted average of all the possible paths that could be taken across the network, with the weights being chosen to be consistent with the laws of electricity and the paths being defined by the disordered structure of the material. This would also be true of any statistic based on path properties of a random walk, and also on the properties associated with mechanical deformation. The form of these statistics are \textit{structurally determined} by the entire disordered structure of each realization.

Statistics which are smoothly defined are ideal for use with the Poisson process, in which there is no interaction between the observed points. For a point process defined by the centres of packed spheres, smoothly defined statistics are often estimated on a length scale which is much smaller than that of the individual sphere diameters. Information about individual sphere shape becomes confounded with interaction effects. This confounding makes the value of the statistic difficult to interpret in terms of features of the realization which can be seen. Structurally determined statistics are determined by the disorder of the model. When physics-based statistics are used on packings, all of the information is captured on the same length scale as the disorder. There is a vast literature associated with the electrical, granular-flow, and fluid-flow models which relates the models to easily-visualized physical phenomena. Structurally-dependent statistics may be more useful than smoothly-defined statistics at summarizing complex structural details that can be identified by eye. 

\subsection{Asymptotics}

Finding the exact distribution of point process statistics is very difficult or impossible, except in the case of the Poisson process. If the realizations were very large, then it may be possible to find asymptotic between-realization distributions for some statistics. 
 
The proofs of central limit theorems and laws of large numbers depend on the individual observations being weakly dependent. If a point process statistic is a smoothly defined average of local measurements, then a central limit theorem can be found as long as the pair correlation function for the point process and some similar higher moments decay to 0 sufficiently fast \cite{penyu}. This result is independent of the nature of the disorder within the realization. If the statistic is not a smoothly defined average, it is not clear that a central limit theorem could be proven. 

Laws of large numbers are of great importance in materials science and physics. If a composite can be modeled by a packing with the spheres and the void having different resistivities, then sufficiently large specimens will have the same bulk resistance when a limit law for specimen resistance exists.  In the case of ideal gases, equilibrium statistical thermodynamics \cite{lls, brush, minlos} derives limit laws from probabilistic models of atomic behaviour. The limiting quantities, experienced as pressure, temperature, volume, and entropy, show no signs of variability arising from the dynamic chaos present at the atomic level. Attempts have been made to create thermodynamic models for the bulk physical properties of granular materials \cite{edwards:1989,edwards:1994}, but these attempts have been based on the assumption that any such model would take the same general form as classical thermodynamics. There is no reason to believe that limit laws for granular materials would resemble those for ideal gases, since the microscale interactions of grains differ from those of atoms and molecules. 

Laws of large numbers are not applicable in many engineering problems. In the case of gases, the limits are taken over collections of more than $10^{15}$ atoms. For granular and composite materials, only thousands or millions of particles will be present, and the interactions among those particles will be much stronger than those found in gases. This suggests that the proper limiting behaviour is not that of convergence to a number, but rather convergence to a random variable. In the absence of theoretical arguments for constructing these limits, limiting distributions for non-average-based statistics will need to be inferred from large samples of model realizations. 

The mathematical intractability of probabilistic limit laws for disordered spatial patterns has resulted in the development of alternative approaches to finding constant large-sample limits for statistics. If a composite is modeled by a lattice packing of spheres, then methods from potential theory can be used to get very good approximations for 
 the bulk electrical conductance \cite{rayleigh,perrins}. When the packing is disordered, these methods  cannot be used on account of the lack of symmetry. Homogenization  \cite{hornung,milton} can be used to find the limiting behaviour of statistics defined by differential equations which have spatially varying boundary conditions. Often, these results are presented as bounds between which the limiting value of the statistic must fall. While the mathematical arguments behind homogenization represent an elegant use of the calculus of variations, the underlying assumptions for these arguments generally cannot be related to the microstructure of any real material. In one case, a mathematical argument has been constructed to establish that a microstructure exists which has the coherent potential approximation for bulk electrical conductance as its limit \cite{milton:1985}.

\section{Inference}

The central problem of inference for sphere packings is model assessment. No proper statistical methods yet exist for determining if a fitted model for a packing has anything beyond a superficial resemblance to the physical system that it is supposed to represent. This has serious implications in physics and engineering, where complex simulation models based on unverifiable assumptions are being used. The history of thermodynamics suggests how methods for assessing fitted models can be developed. 

No algorithm for the simulation of a packing creates realizations in a manner which faithfully represents the physics of packing formation. The DEM models come the closest, but even these are based on assumptions about unobservable mechanisms of energy loss through friction and deformation. Parameters are few in number relative to the size and complexity of the realizations. Parameters often have no physical meaning, especially in the case of rearrangement and ballistic models. 

Model parameters can be fit by methods similar to the method of moments, such as the method of minimum contrast \cite{digrat}. 
This method is based on choosing a set of parameter values which minimizes the difference between a statistic calculated from the model and a statistic calculated from the data. Often, the statistic chosen is the $K-$function. This method ensures that there is some resemblance between the model output and the data, but it says nothing about how far that resemblance can be extended. Since the $K-$function lacks the power to distinguish point patterns which can be distinguished by eye \cite{badsil}, there is no reason to believe that a fit based on it will result in a good fit for other statistics calculated from the packing.

A fitted model must be assessed to determine if the model is useful or if the fit is superficial. Since every realization from the model or from the data may be different, this assessment must be carried out using descriptive statistics which can summarize common aspects of different realizations of the same process and which can identify evidence of important differences between the model and the data. Since there is no guide to which statistics are useful, it is necessary to calculate many different statistics for each realization and then to compare their distributions by means of multivariate statistical methods.  In physical applications, the need for many statistics will require the development of accurate methods for imaging the three-dimensional structure of static and flowing physical packings.

Since no model directly and accurately represents the formation of a physical packing, it is likely that at least one statistic can be found which gives evidence that realizations of a model differ from those of the data. This difference only matters if that statistic has an effect on the physical property being modeled. In a typical application, the packing is being used as part of a model for some specific physical phenomenon such as the electrical conductance of a composite material. If the phenomenon of interest is summarized by a set of response statistics, these statistics can be used to fit the parameters of the model. The only descriptive statistics that are needed for model assessment are those which affect the distribution of the response statistics. In general, there is no way to theoretically determine all of the statistics which can affect the response. Further modeling and testing will be required to establish if statistics which identify a difference between the data and the model also affect the response. 

Some progress has been made towards the development of methods for model assessment in packings. Various methods have been developed for describing deviations from complete spatial randomness in point patterns \cite{cressie}, but it is not clear if these methods will be useful for distinguishing between different types of non-Poisson disorder. Residuals for point processes have been developed \cite{badturn} which can be used to detect spatial trend and interpoint interaction effects. Physicists have used descriptive statistics to quantify differences between simulation models for packings \cite{meak:1993,jm:2000,wouterse} and between different types of physical packings \cite{aste}. Descriptive statistics have also been used to quantify the effects of changing parameters in DEM models \cite{zhang}. These efforts have been used to suggest that some models may be physically wrong, but this condition is too strong. The identification of differences between models is only the beginning of the development of objective and useful model assessment procedures.

\section{Conclusions}

Sphere packings form a very useful class of models for physical phenomena which cannot be modeled formally. They form a class of statistical models consisting entirely of physical and computer simulation procedures which need to be fit to data.

When packings are used as models, the fitted models must be further assessed to determine whether or not the resemblance of the model to the physical system is superficial or useful. If this is not done, there is a severe risk that the fitted model may be misleading as a physical model. The development of assessment procedures requires taking the same approach as early physicists; a specific response must be defined, statistics which affect the response must be identified, and these statistics must be compared for the model and the data. The phenomena being modeled are as well-understood as the properties of gases were in the $18^{th}$ century, and require a similar experimental approach to identify a useful way to summarize everything of importance about them. 

%To be able to define responses and find the statistics which affect them, it is necessary to greatly enlarge the number of descriptive statistics for packings. Mathematical models for physical systems applied out of their physical context will provide a rich source of these statistics, and also have suggested the classification of descriptive statistics as smoothly averaging or structurally defined. 

\section*{Acknowledgments}

I would like to thank Alan Karr of the National Institute of Statistical Sciences and Surendra Shah of the Centre for Advanced Cement-Based Materials for providing the first circumstances in which I found the need to fit sphere packing models to data. Bruce Ankenman, Takeru Igusa, Sanjay Jaiswal, Eric Slud, Dietrich Stoyan, Rolf Turner, and Viqar Husain also provided much useful advice. This work was funded by NISS, the University of Maryland, and NSERC. 

%\bibliography{/Users/jdp/Desktop/spc0.bib}
%\bibliographystyle{unsrt}

\end{document}